\pdfoutput=1

\documentclass[aps,prl,notitlepage,twoside,twocolumn,preprintnumbers,showpacs,showkeys,byrevtex,floatfix,nofootinbib,amsmath,amssymb,superscriptaddress]{revtex4-1}
\usepackage{amscd,array,dsfont,graphicx,mathtools,multirow,rotating,tensor,verbatim,youngtab}
\usepackage[bookmarksnumbered,bookmarksopen=true,bookmarksopenlevel=1,linktocpage,pdfstartview=FitH]{hyperref}
\usepackage[all]{hypcap}

\allowdisplaybreaks[1]

\newcolumntype{M}[1]{>{$}{#1}<{$}}

\DeclareMathOperator{\id}{id}

\DeclareMathOperator{\Or}{O}
\DeclareMathOperator{\SO}{SO}

\DeclareMathOperator{\SL}{SL}

\DeclareMathOperator{\Sp}{Sp}

\newcommand{\half}{\ensuremath{\tfrac{1}{2}}}
\newcommand{\STU}{\ensuremath{STU}}
\newcommand{\SUSY}{\ensuremath{\mathcal{N}}}
\newcommand{\alg}[1]{\ensuremath{\mathfrak{#1}}}
\newcommand{\field}[1]{\ensuremath{\mathds{#1}}}
\newcommand{\rep}[1]{\ensuremath{\mathbf{#1}}}
\newcommand{\tyoung}{\tiny\young}

\newcommand{\ket}[1]{|#1\rangle}

\begin{document}

\title{Four-qubit entanglement from string theory}
\author{L. Borsten}
\email[]{leron.borsten@imperial.ac.uk}
\author{D. Dahanayake}
\email[]{duminda.dahanayake@imperial.ac.uk}
\author{M. J. Duff}
\email[]{m.duff@imperial.ac.uk}
\affiliation{Theoretical Physics, Blackett Laboratory, Imperial College London, London SW7 2AZ, United Kingdom}
\author{A. Marrani}
\email[]{marrani@lnf.infn.it}
\affiliation{Stanford Institute for Theoretical Physics, Stanford University, Stanford, CA 94305-4060, USA}
\author{W. Rubens}
\email[]{william.rubens06@imperial.ac.uk}
\affiliation{Theoretical Physics, Blackett Laboratory, Imperial College London, London SW7 2AZ, United Kingdom}
\setcounter{affil}{1}

\date{\today}

\begin{abstract}
We invoke the black hole/qubit correspondence to derive the classification of four-qubit entanglement. The U-duality orbits resulting from timelike reduction of string theory from $D=4$ to $D=3$ yield 31 entanglement families, which reduce to nine up to permutation of the four qubits.
\end{abstract}

\pacs{11.25.Mj, 03.65.Ud, 04.70.Dy}

\keywords{black hole,  U-duality, qubit, entanglement}

\preprint{Imperial/TP/2010/mjd/2, SU-ITP-10/21}

\maketitle


Recent work has established some intriguing correspondences between two very different areas of theoretical physics: the entanglement of qubits in quantum information theory (QIT) and black holes in string theory. See \cite{Borsten:2008wd} for  a  review. In particular, there is a one-to-one correspondence between the classification of three qubit entanglement \cite {Dur:2000} and the classification of  extremal black holes in the \STU{} supergravity theory \cite{Duff:1995sm,Behrndt:1996hu} that appears in the compactification of string theory from $D=10$ to $D=4$ dimensions.  Moreover,  the Bekenstein-Hawking black hole entropy is provided by the three-way entanglement measure.

The purpose of this paper is to use this black hole/qubit correspondence to address the much more difficult problem of classifying four-qubit entanglement, currently an active area of research in QIT as experimentalists now control entanglement with four qubits \cite{Amselem:2009}.
\begin{table*}[ht]
\caption[Classification of $D=4,\SUSY=2$  \STU{} black holes.]{Various results on four-qubit entanglement\label{tab:authors}}
\begin{ruledtabular}
\begin{tabular*}{\textwidth}{@{\extracolsep{\fill}}clllr@{,}rr@{,}r}
Paradigm                  & Author             & Year & Ref                      & \multicolumn{2}{c}{result mod perms} & \multicolumn{2}{c}{result incl. perms} \\
\hline
\multirow{5}{*}{classes}  & Wallach            & 2004 & \cite{Wallach:2004}      & \multicolumn{2}{c}{?}                & \multicolumn{2}{c}{90}                 \\
                          & Lamata  et al      & 2006 & \cite{Lamata:2006b}      & 8 genuine & 5 degenerate             & 16 genuine       & 18 degenerate       \\
                          & Cao et al          & 2007 & \cite{Cao:2007}          & 8 genuine & 4 degenerate             & 8  genuine       & 15 degenerate       \\
                          & Li et al           & 2007 & \cite{Li:2007c}          & \multicolumn{2}{c}{?}                & $\geq31$ genuine & 18 degenerate       \\
                          & Akhtarshenas et al & 2010 & \cite{Akhtarshenas:2010} & \multicolumn{2}{c}{?}                & 11       genuine & 6  degenerate       \\
\hline
\multirow{3}{*}{families} & Verstraete et al   & 2002 & \cite{Verstraete:2002}   & \multicolumn{2}{c}{9}                & \multicolumn{2}{c}{?}                  \\
                          & Chterental et al   & 2007 & \cite{Chterental:2007}   & \multicolumn{2}{c}{9}                & \multicolumn{2}{c}{?}                  \\
                          & String theory      & 2010 &                          & \multicolumn{2}{c}{9}                & \multicolumn{2}{c}{31}                 \\
\end{tabular*}
\end{ruledtabular}
\end{table*}
Although two and three qubit entanglement is well-understood, the literature on four qubits can be confusing and seemingly contradictory, as illustrated in \autoref{tab:authors}. This is due in part to genuine calculational disagreements, but in part to the use of distinct (but in principle consistent and complementary) perspectives on the criteria for classification. On the one hand there is the ``covariant'' approach which distinguishes the orbits of the equivalence group of Stochastic Local Operations and Classical Communication (SLOCC) by the vanishing or not of covariants/invariants. This philosophy is adopted for the three-qubit case in \cite{Dur:2000,Borsten:2009yb}, for example, where it was shown that three qubits can be tripartite entangled in two inequivalent ways, denoted $W$ and GHZ (Greenberger-Horne-Zeilinger). The analogous four-qubit case was treated, with partial results, in \cite{Briand:2003a}. On the other hand, there is the ``normal form'' approach which considers ``families'' of orbits. Any given state may be transformed into a unique normal form. If the normal form  depends on some of the algebraically independent SLOCC invariants it constitutes a family of orbits parametrized by these invariants. On the other hand a parameter-independent family contains a single orbit. This philosophy is adopted for the four-qubit case
\[
\ket{\Psi}=a_{ABCD}\ket{ABCD}~~A,B,C,D=0,1
\]
in \cite{Verstraete:2002, Chterental:2007}. Up to permutation of the four qubits, these authors found 6 parameter-dependent families called $G_{abcd}$, $L_{abc_2}$, $L_{a_2b_2}$, $L_{a_2 0_{3\oplus \bar{1}}}$, $L_{ab_3}$, $L_{a_4}$ and 3 parameter-independent families called $L_{0_{3\oplus \bar{1}} 0_{3\oplus \bar{1}}}$, $L_{0_{5\oplus \bar{3}}}$, $L_{0_{7\oplus \bar{1}}}$. For example, a family of orbits parametrized by all four of the algebraically independent SLOCC invariants is given by the normal form ${G_{abcd}}$:
\begin{equation}\label{eq:gabcd}
\begin{split}
\frac{a+d}{2}(\ket{0000}+\ket{1111})+\frac{a-d}{2}(\ket{0011}+\ket{1100})\\
+\frac{b+c}{2}(\ket{0101}+\ket{1010})+\frac{b-c}{2}(\ket{1001}+\ket{0110}).
\end{split}
\end{equation}
To illustrate the difference between these two approaches, consider the separable EPR-EPR state $(\ket{00}+\ket{11})\otimes (\ket{00}+\ket{11})$. Since this is obtained by setting $b=c=d=0$ in  \eqref{eq:gabcd} it belongs to the $G_{abcd}$ family, whereas in the covariant approach it forms its own class. Similarly, a totally totally separable $A$-$B$-$C$-$D$ state, such as $\ket{0000}$, for which all covariants/invariants vanish, belongs to the family $L_{abc_2}$, which also contains genuine four-way  entangled states. These interpretational differences were also noted in \cite{Lamata:2006b}.

Our string-theoretic framework lends itself naturally to the ``normal form'' perspective.   We consider $D=4$ supergravity theories in which the moduli parameterize a symmetric space of the form $M_4 = {G_4}/{H_4}$, where $G_4$ is the global U-duality group and $H_4$ is its maximal compact subgroup. After a further  time-like reduction to $D=3$ the moduli space becomes a pseudo-Riemannian symmetric space $M_3^*={G_3}/{H_3^*}$, where $G_3$ is the $D=3$ duality group and $H_3^*$ is a non-compact form of the maximal compact subgroup $H_3$. One finds that geodesic motion on $M_3^*$ corresponds to stationary solutions of the $D=4$ theory \cite{Breitenlohner:1987dg,Gunaydin:2007bg,Bergshoeff:2008be,Bossard:2009we,Bossard:2009at,Levay:2010ua}. These geodesics are parameterized by the Lie algebra valued matrix of Noether charges $Q$ and the problem of classifying the spherically symmetric extremal (non-extremal) black hole solutions consists of classifying the nilpotent (semisimple) orbits of $Q$ (Nilpotent means $Q^n = 0$ for some sufficiently large $n$.)

In the case of the $STU$ model the $D=3$ moduli space $G_3/H_3^*$ is $\SO(4,4)/[\SL(2,\field{R})]^4$ (a para-quaternionic manifold), which yields the Lie algebra decomposition
\begin{equation}\label{eq:4qubitKS}
    \alg{so}(4,4) \cong [\alg{sl}(2,\field{R})]^4 \oplus \rep{(2,2,2,2)}.
\end{equation}
The relevance of  \eqref{eq:4qubitKS} to four qubits was pointed out in \cite{Borsten:2008wd} and recently spelled out more clearly by Levay \cite{Levay:2010ua} who relates four qubits to $D=4$ $STU$ black holes. The Kostant-Sekiguchi correspondence \cite{Collingwood:1993} then implies that the nilpotent orbits of $\SO(4,4)$ acting on the adjoint representation \rep{28}  are in one-to-one correspondence with the nilpotent orbits of $[\SL(2, \field{C})]^4$ acting on the fundamental representation  $\rep{(2,2,2,2)}$ and hence with the classification of four-qubit entanglement.   Note furthermore that it is the complex qubits that appear automatically, thereby relaxing the restriction to real qubits  (sometimes called rebits) that featured in earlier versions of the black hole/qubit correspondence.

Our main result, summarized  in \autoref{tab:realcosets}, is that there are 31 entanglement families which reduce to nine up to permutations of the four qubits. From \autoref{tab:authors} we see that the nine agrees with \cite{Verstraete:2002,Chterental:2007} while the 31 is new. As far as we are aware, the nine four-qubit $[\SL(2, \field{C})]^4$ cosets are also original.

\begin{figure*}[!]
\includegraphics[width=\linewidth]{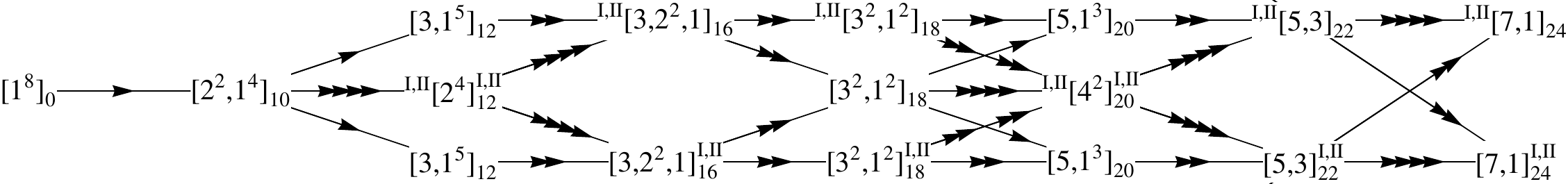}
\caption{$\SO_0(4,4)$ Hasse diagram. The integers inside the bracket indicate the structure of the appropriate Young tableau. The subscript indicates the real dimension of the orbit. The arrows indicate their closure ordering defining a partial order \cite{Djokovic:2000}.\label{fig:hasse}}
\end{figure*}

The nilpotent orbits  required by the Kostant-Sekiguchi theorem are those of $\SO_0(4,4)$, where the $0$ subscript denotes the identity component. These orbits may be labeled by ``signed'' Young tableaux, often referred to as $ab$-diagrams in the mathematics literature. See \cite{Djokovic:2000} and the references therein.   Each signed Young tableau, as listed in \autoref{tab:realcosets}, actually corresponds to a single nilpotent $\Or(4,4)$ orbit of which the $\SO_0(4,4)$ nilpotent orbits are  the connected components. Since $\Or(4,4)$ has four components, for each nilpotent $\Or(4,4)$ orbit there may be either 1, 2 or 4 nilpotent $\SO_0(4,4)$ orbits. This number is also determined by the corresponding signed Young tableau. If the middle sign of every odd length row is  ``$-$'' (``$+$'') there are 2 orbits and we label the diagram to its left (right) with a $I$ or a $II$.  If it only has even length rows there are 4 orbits and we label the diagram to both its left and right with a $I$ or a $II$. If it is none of these it is said to be stable and there is only one orbit. The signed Young tableaux together with their labellings, as listed in \autoref{tab:realcosets}, give a total of 31 nilpotent $\SO_0(4,4)$ orbits, which are  summarized in \autoref{fig:hasse}.
We also supply the complete list of the associated cosets in \autoref{tab:realcosets}, some of which may be found in \cite{Bossard:2009we}.

The $STU$ model describes $\SUSY=2$ supergravity coupled to three vector multiplets and the Hawking temperature and Bekenstein-Hawking entropy of the $STU$ black holes will depend on their mass and a maximum of 8 charges (four electric and four magnetic). Through scalar-dressing, these charges can be grouped into the $\SUSY=2$ central charge $z$ and three ``matter charges'' $z_a$ ($a=1, 2, 3$), which exhibit a triality (corresponding to permutation of three of the qubits). The black holes are divided into extremal or non-extremal according as the temperature is zero or not. The orbits are nilpotent or semisimple, respectively. Depending on the values of the charges, the extremal black holes are further divided into small or large according as the entropy is zero or not. The small ones are termed lightlike, critical or doubly critical according as the minimal number of representative electric or magnetic charges is 3, 2 or 1. The lightlike case is split into one 1/2-BPS solution, where the charges satisfy $z_1=0, |z|^2=4|z_2|^2=4|z_3|^2$ and three non-BPS solutions, where the central charges satisfy $z=0, |z_1|^2=4|z_2|^2=4|z_3|^2$ or $z_2=0, |z_3|^2=4|z_1|^2=4|z|^2$ or $z_3=0, |z_2|^2=4|z_1|^2=4|z|^2$. The critical case splits into three 1/2-BPS solutions with  $z=z_a\not=0, z_b=z_c=0$ and three non-BPS cases with $z=z_a=0, z_b=z_c\not=0$, where $a\not=b\not=c$. The doubly critical case is always 1/2-BPS with $|z|^2=|z_1|^2=|z_2|^2=|z_3|^2$ and vanishing sum of the $z_a$ phases. The large black holes  may also be 1/2-BPS or non-BPS. One subtlety is that some extremal cases,  termed ``extremal'', cannot be obtained as limits of non-extremal black holes. The matching of the extremal classes to the nilpotent orbits is given in \autoref{tab:realcosets}.

\begingroup
\squeezetable
\newlength\BHCol
\setlength\BHCol{1.4cm}
\newlength\RepCol
\setlength\RepCol{2.4cm}
\begin{table*}[!]
\caption{Each black hole nilpotent $\SO_0(4,4)$ orbit  corresponds to a 4-qubit nilpotent $[\SL(2,\field{C})]^4$ orbit. $z_H$ is the horizon value of the $\SUSY=2$, $D=4$ central charge
.\label{tab:realcosets}}
\begin{ruledtabular}
\begin{tabular*}{\textwidth}{@{\extracolsep{\fill}}>{\centering}m{\BHCol}*{4}{M{c}}>{\centering$}m{\RepCol}<{$}cM{c}}
\multicolumn{3}{c}{$STU$ black holes}                                                                                                                                                                                                                                                                          & \multirow{2}{*}{$\dim_\field{R}$} & \multicolumn{4}{c}{Four qubits}                                                                                                                                                                                                   \\
\cline{1-3}\cline{5-8}\\[-6pt]
description                           & \text{Young tableaux}                                                                                                                       & \SO_0(4,4)\text{ coset}                                                                                                  &                                   & [\SL(2,\field{C})]^4\text{ coset}                                                                        & \text{nilpotent rep}                                            &  	     & \text{family}                               \\
\\[-6pt]\hline\hline\\[-6pt]
trivial                               & \text{trivial}                                                                                                                              & \frac{\SO_0(4,4)}{\SO_0(4,4)}                                                                                            & 1                                 & \frac{[\SL(2,\field{C})]^4}{[\SL(2,\field{C})]^4}                                                        & 0                                                               & $\in$  & G_{abcd}                                    \\
\\[-6pt]\hline\\[-6pt]
doubly-critical \half BPS             & \tyoung(+-,-+,+,-,+,-)                                                                                                                      & \frac{\SO_0(4,4)}{[\SL(2,\field{R})\times \SO(2,2,\field{R})]\ltimes[(\rep{2,4})^{(1)} \oplus{\rep1}^{(2)}]}             & 10                                & \frac{[\SL(2,\field{C})]^4}{[\SO(2,\field{C})]^3\ltimes\field{C}^4}                                      & \ket{0110}                                                      & $\in$  & L_{abc_2}                                   \\
\\[-6pt]\hline\\[-6pt]
                                      & \tyoung(-+-,+,-,+,-,+)                                                                                                                      & \frac{\SO_0(4,4)}{\SO(3,2;\field{R})\ltimes[(\rep{5\oplus1})^{(2)}]}                                                     &                                   &                                                                                                          &                                                                 &        &                                             \\
\\[-6pt]
critical, \half BPS and non-BPS       & \tyoung(+-+,-,+,-,+,-)                                                                                                                      & \frac{\SO_0(4,4)}{\SO(2,3;\field{R})\ltimes[(\rep{5\oplus1})^{(2)}]}                                                     & 12                                & \frac{[\SL(2,\field{C})]^4}{[\SO(3,\field{C})\times \field{C}]\times[\SO(2,\field{C})\ltimes \field{C}]} & \ket{0110}+\ket{0011}                                           & $\in$  & L_{a_2b_2}                                  \\
\\[-6pt]
                                      & \begin{pmatrix}I,II&\tyoung(+-,-+,+-,-+)&I,II\end{pmatrix}                                                                                  & \frac{\SO_0(4,4)}{\Sp(4,\field{R})\ltimes[(\rep{5\oplus1})^{(2)}]}                                                       &                                   &                                                                                                          &                                                                 &        &                                             \\
\\[-6pt]\hline\\[-6pt]
lightlike \half BPS and non-BPS       & \begin{array}{c}\begin{pmatrix}I,II&\tyoung(+-+,-+,+-,-)\;\end{pmatrix}\\\begin{pmatrix}\;\tyoung(-+-,+-,-+,+)&I,II\end{pmatrix}\end{array} & \frac{\SO_0(4,4)}{\SL(2,\field{R})\ltimes[(2\times {\rep 2})^{(1)}\oplus(3\times{\rep 1})^{(2)}\oplus{\rep 2}^{(3)}]}    & 16                                & \frac{[\SL(2,\field{C})]^4}{[\SO(2,\field{C})\ltimes\field{C}]\times \field{C}^2}                        & \ket{0110}+\ket{0101}+\ket{0011}                                & $\in$  & L_{a_2 0_{3\oplus \bar{1}}}                 \\
\\[-6pt]\hline\\[-6pt]
large non-BPS $z_H \neq 0$            & \tyoung(-+-,+-+,-,+)                                                                                                                        & \frac{\SO_0(4,4)}{\SO(1,1,\field{R})\times\SO(1,1,\field{R})\ltimes[(\rep{(2,2)\oplus(3,1)})^{(2)}\oplus{\rep 1}^{(4)}]} & 18                                & \frac{[\SL(2,\field{C})]^4}{\field{C}^3}                                                                 & \tfrac{i}{\sqrt{2}}(\ket{0001}+\ket{0010}-\ket{0111}-\ket{1011})& $\in$  & L_{ab_3}                                    \\
\\[-6pt]\hline\\[-6pt]
                                      & \tyoung(-+-+-,+,-,+)                                                                                                                        & \frac{\SO_0(4,4)}{\SO(2,1;\field{R})\ltimes[\rep{1}^{(2)}\oplus\rep{3}^{(4)}\oplus\rep{1}^{(6)}]}                        &                                   &                                                                                                          &                                                                 &        &                                             \\
\\[-6pt]
``extremal''                          & \tyoung(+-+-+,-,+,-)                                                                                                                        & \frac{\SO_0(4,4)}{\SO(1,2;\field{R})\ltimes[\rep{1}^{(2)}\oplus\rep{3}^{(4)}\oplus\rep{1}^{(6)}]}                        & 20                                & \frac{[\SL(2,\field{C})]^4}{\SO(2,\field{C})\times\field{C}}                                             & i\ket{0001}+\ket{0110}-i\ket{1011}                              & $\in$  & L_{a_4}                                     \\
\\[-6pt]
                                      & \begin{pmatrix}I,II&\tyoung(+-+-,-+-+)& I,II\end{pmatrix}                                                                                   & \frac{\SO_0(4,4)}{\Sp(2,\field{R})\ltimes[\rep{1}^{(2)}\oplus\rep{3}^{(4)}\oplus\rep{1}^{(6)} ]}                         &                                   &                                                                                                          &                                                                 &        &                                             \\
\\[-6pt]\hline\hline\\[-6pt]
large \half BPS and non-BPS  $z_H=0$  & \begin{array}{c}\begin{pmatrix}I,II&\tyoung(+-+,+-+,-,-)\end{pmatrix}\\\begin{pmatrix}\;\tyoung(-+-,-+-,+,+)&I,II\end{pmatrix}\end{array}   & \frac{\SO_0(4,4)}{\SO(2,\field{R})\times\SO(2,\field{R})\ltimes[(\rep{(2,2)\oplus(3,1)})^{(2)}\oplus{\rep 1}^{(4)}]}     & 18                                & \frac{[\SL(2,\field{C})]^4}{[\SO(2,\field{C})]^2\times \field{C}}                                        & \ket{0000}+\ket{0111}                                           & $\in$  & L_{0_{3\oplus \bar{1}} 0_{3\oplus \bar{1}}} \\
\\[-6pt]\hline\\[-6pt]
``extremal''                          & \begin{array}{c}\begin{pmatrix}I,II&\tyoung(-+-+-,+-+)\end{pmatrix}\\\begin{pmatrix}\;\tyoung(+-+-+,-+-)&I,II\end{pmatrix}\end{array}       & \frac{\SO_0(4,4)}{\field{R}^{3(2)}\oplus\field{R}^{1(4)}\oplus\field{R}^{2(6)}}                                          & 22                                & \frac{[\SL(2,\field{C})]^4}{\field{C}}                                                                   & \ket{0000}+\ket{0101}+\ket{1000}+\ket{1110}                     & $\in$  & L_{0_{5\oplus \bar{3}}}                     \\
\\[-6pt]\hline\\[-6pt]
``extremal''                          & \begin{array}{c}\begin{pmatrix}I,II&\tyoung(+-+-+-+,-)\end{pmatrix}\\\begin{pmatrix}\;\tyoung(-+-+-+-,+)&I,II\end{pmatrix}\end{array}       & \frac{\SO_0(4,4)}{\field{R}^{(2)}\oplus\field{R}^{2(6)}\oplus\field{R}^{(10)}}                                           & 24                                & \frac{[\SL(2,\field{C})]^4}{\id}                                                                         & \ket{0000}+\ket{1011}+\ket{1101}+\ket{1110}                     & $\in$  & L_{0_{7\oplus \bar{1}}}                     \\
\end{tabular*}
\end{ruledtabular}
\end{table*}
\endgroup


It follows from the Kostant-Sekiguchi theorem that there are 31 nilpotent orbits for the SLOCC-equivalence group acting on the representation space of four qubits. For each nilpotent orbit there is precisely one family of SLOCC orbits since each family contains one nilpotent orbit on setting all invariants to zero. The nilpotent orbits and their associated families are summarized in \autoref{tab:realcosets}, which is split into upper and lower sections according as  the nilpotent orbits belong to parameter-dependent or parameter-independent families.

If one allows for the permutation of the four qubits the connected components of each $\Or(4,4)$ orbit are re-identified reducing the count to 17. Moreover, these 17 are further grouped under this permutation symmetry into just nine nilpotent orbits. It is not difficult to show that these nine cosets match the nine families of \cite{Verstraete:2002,Chterental:2007}, as listed in the final column of \autoref{tab:realcosets} (provided we adopt the version of $L_{ab_3}$ presented in \cite{Chterental:2007} rather than in \cite{Verstraete:2002}). For example, the state representative $L_{0_{3\oplus \bar{1}}0_{3\oplus \bar{1}}}$
\begin{equation}\label{eq:l053}
\begin{split}
\ket{0111}+\ket{0000}
\end{split}
\end{equation}
is left invariant by the $[\SO(2,\field{C})]^2\times \field{C}$	 subgroup, where $[\SO(2,\field{C})]^2$ is the stabilizer of the three-qubit GHZ state \cite{Borsten:2009yb}. In contrast, the four-way entangled family $L_{0_{7\oplus \bar{1}}}$, which is the ``principal'' nilpotent orbit \cite{Collingwood:1993}, is not left invariant by any subgroup.
Note that the total of 31 does not follow trivially by permuting the qubits in these nine. Naive permutation produces far more than 31 candidates which then have to be reduced to SLOCC inequivalent  families.

There is a satisfying consistency of this process with respect to the covariant approach. For example, the covariant classification has four biseparable classes $A$-GHZ, $B$-GHZ, $C$-GHZ and $D$-GHZ which are then identified as a single class under the permutation symmetry. These four classes are in fact the four nilpotent orbits corresponding to the families $L_{0_{3\oplus \bar{1}} 0_{3\oplus \bar{1}}}$ in \autoref{tab:realcosets}, which are also identified as a single nilpotent orbit under permutations. Similarly, each of the four $A$-W classes is a nilpotent orbit belonging to one of the four families labeled $L_{a_2 0_{3\oplus \bar{1}}}$ which are again identified under permutations. A less trivial example is given by the six $A$-$B$-EPR classes of the covariant classification. These all lie in the single family $L_{a_2b_2}$ of \cite{Verstraete:2002}, which is defined up to permutation. Consulting \autoref{tab:realcosets} we see that, when not allowing permutations, this family splits into six pieces, each containing one of the six $A$-$B$-EPR classes. Finally, the single totally separable class $A$-$B$-$C$-$D$ is the single nilpotent orbit inside the single family $L_{abc_2}$ which maps into itself under permutations.


Falsifiable predictions in the fields of high-energy physics or cosmology are hard to come by, especially for ambitious attempts, such as string/M-theory, to accommodate all the fundamental interactions. In the field of quantum information theory, however, previous work has shown that the stringy black hole/qubit correspondence can reproduce well-known results in the classification of two and three qubit entanglement. In this paper this correspondence has been taken one step further to predict new results in the less well-understood case of four-qubit entanglement that can in principle be tested in the laboratory.

\begin{acknowledgements}
This work was supported in part by the STFC under rolling Grant No. ST/G000743/1.  The work of A.M. has been supported by an INFN visiting Theoretical Fellowship at SITP, Stanford University, Stanford, CA, USA. This work was completed at the CERN theory division, supported by ERC Advanced Grant ``Superfields''. We are grateful to Sergio Ferrara for useful discussions and for his hospitality. D.D. is grateful to Steven Johnston for useful discussions.
\end{acknowledgements}

%

\end{document}